\documentclass[aps,prl,twocolumn,superscriptaddress,showpacs]{revtex4-1}
\usepackage{graphicx}
\usepackage{dcolumn} 
\usepackage{bm}      
\usepackage[colorlinks]{hyperref}

\begin{document}

This manuscript has been authored by UT-Battelle,
LLC under Contract No. DE-AC05-00OR22725 with
the U.S. Department of Energy. The United States
Government retains and the publisher, by accepting
the article for publication, acknowledges that the
United States Government retains a non-exclusive, paidup,
irrevocable, world-wide license to publish or reproduce
the published form of this manuscript, or
allow others to do so, for United States Government
purposes. The Department of Energy will
provide public access to these results of federally
sponsored research in accordance with the DOE
Public Access Plan(http://energy.gov/downloads/doepublic-access-plan).
\clearpage

\title{Low-temperature crystal and magnetic structure of $\alpha$-RuCl$_3$ }

\author{H.~B.~Cao}
\email{caoh@ornl.gov}\affiliation{Quantum Condensed Matter Division, Oak Ridge National Laboratory, Oak Ridge, TN 37831, USA}

\author{A.~Banerjee}
\email{banerjeea@ornl.gov}\affiliation{Quantum Condensed Matter Division, Oak Ridge National Laboratory, Oak Ridge, TN 37831, USA}

\author{J.-Q.~Yan}
\affiliation{Materials Science and Technology Division, Oak Ridge National Laboratory, Oak Ridge, TN 37831, USA}
\affiliation{Department of Materials Science and Engineering, University of Tennessee, Knoxville, TN 37996, USA}

\author{C.~A.~Bridges}
\affiliation{Chemical Sciences Division, Oak Ridge National Laboratory, Oak Ridge, TN 37831, USA}

\author{M.~D.~Lumsden}
\affiliation{Quantum Condensed Matter Division, Oak Ridge National Laboratory, Oak Ridge, TN 37831, USA}

\author{ D.~G. Mandrus}
\affiliation{Materials Science and Technology Division, Oak Ridge National Laboratory, Oak Ridge, TN 37831, USA}
\affiliation{Department of Materials Science and Engineering, University of Tennessee, Knoxville, TN 37996, USA}

\author{D.~A.~Tennant}
\affiliation{Neutron Sciences Directorate, Oak Ridge National Laboratory, Oak Ridge, TN 37831, USA}

\author{B.~C.~Chakoumakos}
\affiliation{Quantum Condensed Matter Division, Oak Ridge National Laboratory, Oak Ridge, TN 37831, USA}

\author{ S.~E.~Nagler}
\affiliation{Quantum Condensed Matter Division, Oak Ridge National Laboratory, Oak Ridge, TN 37831, USA}
\affiliation{Bredesen Center,   University of Tennessee,  Knoxville,  TN  37996, USA }

\date{\today}

\begin{abstract}

Single crystals of the Kitaev spin-liquid candidate $\alpha$-RuCl$_3$ have been studied to determine low-temperature bulk properties, structure and the magnetic ground state. Refinements of x-ray diffraction data show that the low temperature crystal structure is described by space group $C2/m$ with a nearly-perfect honeycomb lattice exhibiting less than 0.2 \% in-plane distortion. The as-grown single crystals exhibit only one sharp magnetic transition at $T_{N}$ = 7~K.  The magnetic order below this temperature exhibits a propagation vector of $k$ = (0, 1, 1/3), which coincides with a 3-layer stacking of the $C2/m$ unit cells.  Magnetic transitions at higher temperatures up to 14~K can be introduced by deformations of the crystal that result in regions in the crystal with a 2-layer stacking sequence.   The best fit symmetry allowed magnetic structure of the as-grown crystals shows that the spins lie in the $ac$-plane, with a zigzag configuration in each honeycomb layer.  The three layer repeat out-of-plane structure can be refined as a 120$^o$ spiral order or a collinear structure with spin direction 35$^o$ away from the $a$-axis. The collinear spin configuration yields a slightly better fit and also is physically preferred. The average ordered moment in either structure is less than 0.45(5) $\mu_B$ per Ru$^{3+}$ ion.

\end{abstract}



\maketitle

\section{I. Introduction} The Kitaev Quantum Spin Liquid  (QSL) is an exotic state of matter  that exhibits fractionalized excitations in the form of Majorana fermions, and a realization of this physics has potential applications in the field of quantum information \cite{kitaev06, nayak08,yamashita10, han12, knolle14}.   The state can be realized by considering a honeycomb lattice of $S$=1/2 moments with either ferromagnetic or antiferromagnetic bond directional Ising interactions, also referred to as Kitaev interactions.  The necessary terms in an effective spin Hamiltonian can be realized in practice by octahedrally coordinated low-spin $d^{5}$ magnetic ions  with strong spin-orbit coupling \cite{jackeli09}. It is highly desirable to investigate the magnetic structure and excitations of these materials using neutron scattering techniques. Much attention has been paid to iridates containing  honeycomb lattices with Ir$^{4+}$ ions  \cite{singh10,biffin14, coldea14, singh12, choi12, ye12, chaloupka13, rau15, chun15} as candidate materials for Kitaev physics. The challenges presented by the  large neutron absorption cross-section and rapid decrease of the Ir magnetic form factor have limited the neutron studies of honeycomb iridates to date \cite{coldea14, choi12, ye12}. On the other hand, it is much more feasible to carry out detailed neutron scattering investigations of the binary halide $\alpha$-RuCl$_3$, which has been proposed as an alternate candidate to realize Kitaev physics\cite{sears15, sandilands15, majumder15,kubota15, kobayashi92, arnab15, radu15}. In this material, honeycomb layers of octahedrally coordinated low spin Ru$^{3+}$ ions  are coupled  via weak van der Waals inter-layer interactions \cite{str1, str2} (see Fig.~1).  The single ion ground state is confirmed as effective $J$=1/2 \cite{kobayashi92,majumder15, arnab15,radu15} and interestingly both inelastic neutron scattering \cite{arnab15} and Raman scattering \cite{sandilands15}  show evidence for continuum magnetic response that could be associated with fractionalized excitations.  Neutron diffraction studies \cite{sears15,arnab15,radu15} have shown that the materials order in-plane with a zigzag spin structure which can arise from the Kitaev-Heisenberg model \cite{chaloupka10} as well as Kitaev-dominated models with additional more complicated interactions\cite{rau14, chaloupka15, rous15}. In the course of investigations of $\alpha$-RuCl$_3$, it has become clear that sample-dependent issues have led to some confusion.

With hexagonal or honeycomb layers that are weakly coupled, it is extremely easy for the system to form stacking faults \cite{radu15,kim15}, and in fact various polytypes characterized by different stacking sequences \cite{smithyoder56} have nearly the same energy.   This may be responsible for the fact that different crystal structures  have been reported, including both the trigonal (T) space group $P3_112$ where each unit cell contains three honeycomb layers\cite{str0,str1,str2},  and the one layer per unit cell monoclinic (M) space group $C2/m$ \cite{strmono}.  The low temperature magnetic ordering transitions have similarly shown some discrepancies.   Specific heat and susceptibility measurements on powders \cite{majumder15,kobayashi92, arnab15,radu15} have consistently shown a broad N\'eel transition at 14~K, however many previously investigated single-crystals show evidence for  two separate N\'eel transitions at $T_N$  = 8~K and 14~K, and perhaps others \cite{arnab15}, with some suggestions that these are associated with exotic physics \cite{sears15,kubota15}.   A previous neutron diffraction study of single crystals with stacking faults indicated by a broad mosaicity in the direction perpendicular to the honeycomb planes \cite{arnab15}  showed (using T  nomenclature) that the 14~K  transition  is associated with  a (1/2, 0, 3/2)$_T$ ordering wave vector corresponding  to a bilayer ABAB  antiferromagnetic stacking, while the 8~K transition is associated with  a (1/2, 0, 1)$_T$ wave vector indicating ABC periodicity (see the transformation matrix in the note \cite{matrix}). From here on we adopt the $C2/m$ symmetry convention, and henceforth reflections mentioned in this paper are indexed in the monoclinic lattice except for those with the label ”T”.  These structures coexist in physically separated domains in the material in different proportions in different crystals.   As mentioned above, stacking faults are ubiquitous in these layered materials and in diffraction patterns show up as diffuse rods. The dependence of the magnetic wave vectors on the stacking sequence greatly complicates the solution of the crystal and magnetic structures.  A recent report of structural and magnetic results on crystals that were untwinned but exhibited significant stacking faults found that the materials were monoclinic and apparently exhibited a single, broad magnetic transition near 14~K \cite{radu15}.
Given that the magnetic properties can be dependent on the stacking sequence, it is crucial to study and refine the structures on single crystals of $\alpha$-RuCl$_3$ with minimal stacking faults.  Here we report the results of studies of such crystals.  X-ray refinements of the low temperature structure yield a monoclinic $C 2/m$ space group with a symmetrical in-plane honeycomb lattice.  Consistently, bulk characterization shows that the as-grown crystals exhibit a single well-defined and sharp magnetic transition at 7~K, with no signature of order apparent at 14~K.  Mechanically deforming the crystals induces a magnetic transition at 14~K and the crystals subject to repeated deformations eventually exhibit only one broad transition at 14~K.
The 14~K transition with its observed two-layer periodicity is likely associated with a high density of stacking faults. Neutron diffraction on a large high-quality single crystal conforms to the same monoclinic structure and only one transition at 7~K corresponding to an in-plane zigzag structure with 3-layer periodic magnetic ground state. The ordered moment is found to be small with the spin direction confined to the $ac$-plane. Two symmetry-allowed magnetic models have been found to be consistent with the observed data: one is collinear and the other exhibits an out-of-plane spiral. The best overall refinement has been obtained with the collinear model and leads to moment directions pointing 35$^o$ away from the $a$-axis.  The possible significance of this is discussed below.

\section{II. Experimental details} Commercial RuCl$_3$ powder (which has Ru$^{4+}$ impurities in the form of partial oxides like RuOCl$_2$ and Ru$_2$OCl$_6$ and some unreacted Ru metal) \cite{hyde65} was purified and  used to grow single crystals by vapor transport techniques at ORNL.  A careful purification has been found to be critical for the growth of consistently high quality and sizeable single crystals. The as-grown crystals are shiny plates with a typical thickness of 0.5 mm but with some as large as 15 x 20 x 2 mm$^3$. The temperature dependent specific heat was measured using a 9\,T Quantum Design Physical Property Measurement System. For single crystal x-ray diffraction, a 100x100x30 $\mu$m$^3$ as-grown untwinned crystal was carefully suspended in paratone oil and mounted on a plastic loop attached to copper pin/goniometer. Single crystal x-ray diffraction data reported were collected with molybdenum K$_\alpha$ radiation ($\lambda$ = 0.71073 \AA) using a Rigaku XtaLAB PRO diffractometer  equipped with  a Dectris Pilatus 200K detector and an Oxford N-HeliX cryocooler.   The raw data were examined using MAX3D \cite{max3d}. Peak indexing and integration were done using 'd*trek' in the CrystalClear package \cite{rigaku05}. A numerical absorption correction was applied using ABSCOR \cite{higashi}. The SIR-2011 in WinGX and SHELXL-2013 software packages was used for data processing and structure solution and refinement\cite{sheldrick, farrugia, burla}. Crystal structure projections were made with VESTA\cite{momma}. Single crystal neutron diffraction was performed at the HB-3A Four-circle Diffractometer equipped with a 2D detector at the High Flux Isotope Reactor(HFIR) at ORNL. Neutron wavelengths of 1.005~\AA~ and 1.546~\AA~were used from a bent perfect Si-331 and Si-220 monochromator \cite{hb3a}. Neutrons with a short wavelength of 1.005 \AA~enabled a large reciprocal space coverage and were used to measure the nuclear structure.  The magnetic reflections at 4.2~K and the temperature dependent magnetic intensity were measured using the longer wavelength of 1.546 \AA~which also had the benefit of higher flux on sample. The nuclear and magnetic structure refinements and representation analysis were carried out with the FullProf Suite\cite{fullprof}.

\begin{figure}[tbp]
\linespread{1}
\par
\begin{center}
\includegraphics[width=3.7in]{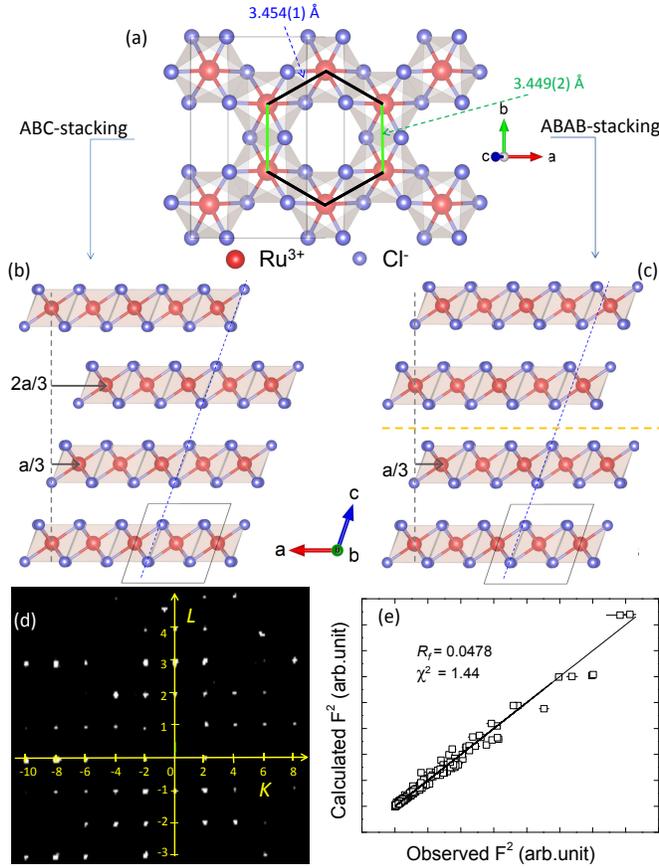}
\end{center}
\par
\caption{(Color Online) (a) The in-plane structure of layered compound $\alpha$-RuCl$_3$ viewed perpendicular to the layers along the $c^*$ axis showing the positions of the Ru$^{3+}$ ions (red) and Cl$^{-}$ ions (blue) in $C2/m$ space group symmetry. Two types of Ru$^{3+}$$-$Ru$^{3+}$ distances are highlighted with blue (3.454(1)\AA) and green (3.449(2) \AA) lines  (b) ABC-stacking:   the out-of-plane structure viewed along the $b$ axis, showing the layered structure as well as the octahedra tilted along $c$. Every layer is translated by $a$/3, such that every fourth layer falls on top of the first  layer when viewed along $c^*$ (dotted line). The blue dashed line along the $c$-axis indicates the perfect stacking for this ABC-stacking. Details of the monoclinic structure are in the text as well as in Table 1. (c) ABAB-stacking viewed in the same orientation as that in (b), with the top two layers twinned 120$^o$ with respect to the bottom two layers. The neighboring layer is translated back and forth by $a$/3, guided by the dotted grey line (details in text). The blue dashed line along the $c$-axis indicates the ABAB stacking is faulty stacking (orange dashed line showing the layer at which the fault occurs). (d) [$0KL$] cut of the reciprocal space measured by single crystal x-ray diffraction at 60~K. (e) the observed squared structure factors versus the calculated ones from the single crystal x-ray data refinement. }
\end{figure}

\begin{figure}[tbp]
\linespread{1}
\par
\begin{center}
\includegraphics[width=3.5 in]{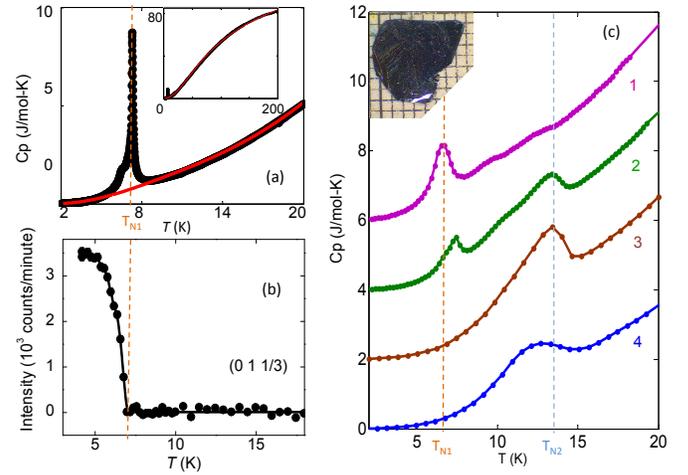}
\end{center}
\par
\caption{(Color Online) (a) The heat capacity of an as-grown single crystal of $\alpha$-RuCl$_3$ from 2-20~K shows the low-temperature region exhibiting one sharp N\'eel transition at $T_{N1}$ = 7~K. As determined by the neutron diffraction reported here the in-plane magnetic order is zig-zag with out of plane order corresponding to ABC stacking of the layers. The inset shows the data in the large temperature range of 2-200~K. The data excluding 5-10~K transition region can be fit to a 2D Debye expression $C_p (T)=ANk(\frac{T} {\theta_D})^2\int_0^{\frac{\theta_D}{T}}\frac{x^2}{e^x-1}dx$ with $\theta_D$=209(2)~K, similar to powder \cite{arnab15}.  (b) The thermal evolution of the intensity of the magnetic peak (0, 1, 1/3)(trigonal (1/2, 0, 1)$_T$) as measured at HB-3A shows a rapid fall-off near  $T_{N1}$ =7~K.  A crystal from the same batch as shown in (c)-inset with similar dimensions was used for the single crystal neutron diffraction at HB-3A.  (c)The heat capacity data on the same single crystal when subject to artificial deformation. The violet curve-1 represents the marginally deformed crystal and shows a rather suppressed anomaly at $T_{N1}$ as compared to an as-grown crystal. When this crystal is manually deformed, the green curve-2 is obtained.  Additional transitions, most prominently at $T_{N2}$ =14~K,  appear, while the anomaly at the $T_{N1}$ transition loses strength.  When deformed further, the anomaly at $T_{N1}$ completely vanishes leaving behind a broad anomaly $T_{N2}$ as the only strong feature (brown curve-3).  The heat-capacity of the deformed crystal resembles that of powder $\alpha$-RuCl$_3$ (blue curve-4). The inset is a picture of the as-grown crystal against a grid of 1~mm$^2$. }
\end{figure}

\begin{figure} [tbp]
\par
\begin{center}
\includegraphics[width=3.6 in]{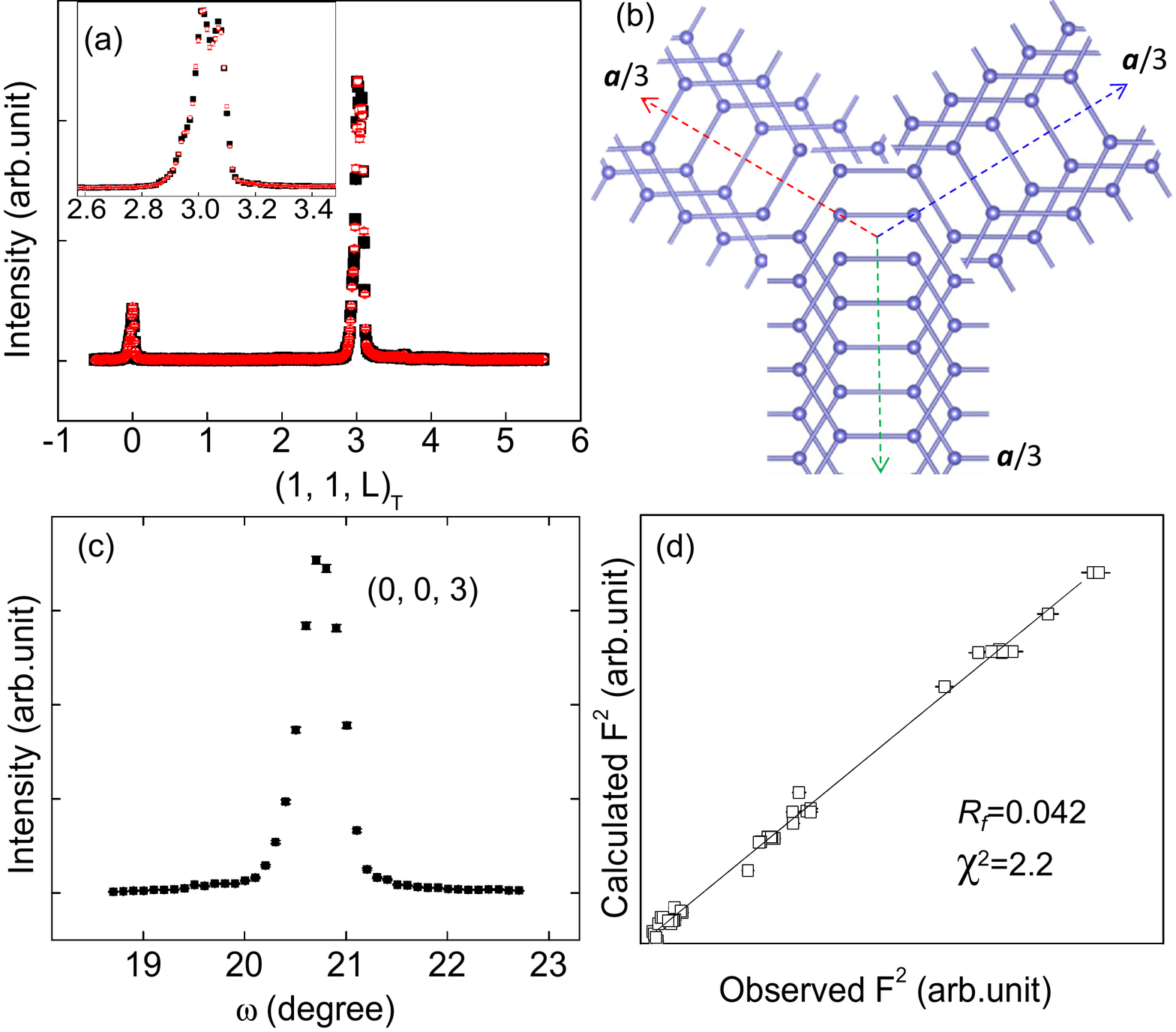}
\end{center}
\par
\caption{(Color Online) (a) L-scans of (1, 1, L)$_T$, (2, 0, L/3) in monoclinic notation, measured by single crystal neutron diffraction at 5~K (black solid squares) and 10~K (red open circles) at HB-3A. The reflections can be indexed in a trigonal lattice but the intensities are incompatible with a trigonal lattice symmetry.  No structural change was observed from 5~K to 10~K. The inset shows a close-up at (1, 1, 3)$_T$. (b)  Three monoclinic structure domains form in the crystal measured by neutron diffraction. Each domain has the honeycomb layer shifted by $a$/3 along the direction perpendicular to one of honeycomb edges from one layer to the next, which yields a pseudo trigonal pattern in the neutron diffraction measurement. (c) Rocking curve scan at (0, 0, 3) measured by neutron diffraction exhibits a resolution limited width of ~0.5$^o$ along the stacking direction, indicating a high quality crystal with only a few stacking faults. (d) Observed squared structure factors versus the calculated ones from the neutron data refinement.}
\end{figure}

\section{III. Results}
\subsection{III-a X-ray diffraction }

More than 1300 reflections (501 unique reflections) were collected and used in the structure refinements for each temperature by XRD on a single crystal (0.003~mm$^3$). To assess the signal to background, we note that in a time interval producing 24000 counts at the Bragg peak (1, 3, 1), the signal registered at a background location varied from 0-2 counts. Given this signal-to-noise, at all temperatures our pattern showed no diffuse scattering intensity at the $K$ = 3$n$ + 1 or 3$n$ + 2 ($n$ = integer) locations.  At these locations diffuse scattering rods along $L$ are characteristic of occasional $\pm$$b$/3 shifts in the $ab$-plane, and have been referred to by Johnson $et$~$al$. as stacking faults of type ``a'' \cite{radu15}. In fact, at room temperature, and also at 250~K, our diffraction pattern was free from any observable diffuse scattering. Upon cooling to 100~K, the pattern shows a faint diffuse scattering present along the $L$-direction in the reflections
 ($H$,$K$,$L$) with $K$ = 3$n$. The appearance of these new diffuse scattering rods at low temperature is qualitatively similar to the observations of type ``b'' stacking fault scattering reported in Ref.\cite{radu15}.  This diffuse scattering does not appear as a continuous rod of scattering but rather as an array of weak broad peaks at the $L$ = $n$ $\pm$1/3 locations, with intensity varying between 10-15 counts at the brightest spots. The upper bound for the peak intensity of diffuse scattering was 5x10$^{-4}$ times the most intense Bragg peaks. The tiny diffuse scattering signifies that our samples have an extremely low density of stacking faults and also allows our structural refinement to use all the peaks. The ($0KL$) slice of the diffraction data (Fig.~1d) shows a typical slice of the XRD data. The structure solution determined a monoclinic symmetry of $C2/m$ at 60~K with lattice parameters $a$=5.981(2)~\AA, $b$=10.354(4)~\AA, $c$=6.014(2)~\AA, and $\beta$=108.800(1)$^o$.  Table 1 lists
the detailed structure parameters.  Figure 1 shows the lattice structure in different views.  Each monoclinic unit cell contains one honeycomb layer. The three Ru-Ru distances in the honeycomb are 3.454(2)~\AA, 3.454(2)~\AA~and 3.448(2)~\AA~(along $b$), indicating a distortion of less than 0.2\% (Fig.~1a). Three Ru-Cl-Ru bond angles are also similar, with values of 93.9(1)$^o$, 93.9(1)$^o$ and 93.6(2)$^o$. This indicates a nearly perfect honeycomb lattice, with a honeycomb distortion much lower than previously reported \cite{radu15}.

The data fit quality is shown in Fig.~1e by plotting the observed squared structure factors versus the calculated ones. With all collected reflections, the refinement yields a goodness of fit 1.44, which further indicates the diffuse scattering has a negligible effect on the refinement.  In this regard, the crystals studied here are different from those reported by Johnson $et$~$al.$ \cite{radu15} where strong diffuse scattering of types ``a'' and ``b'' had a significant effect on the structure refinement.  In addition, for crystals with an edge length below 100 $\mu$m, the ratio of the crystal surface (assume the surface contains three honeycomb layers) to the crystal volume is around 10$^{-4}$, suggesting that the diffuse scattering observed in the present experiment may arise primarily from the crystal surface region.  A total absence of the diffuse scattering in the neutron diffraction pattern in thicker crystals supports this conjecture.

\begin{widetext}

\begin{table}

\caption{The structure parameters of $\alpha$-RuCl$_3$ measured at 60~K by single crystal x-ray diffraction. The space group is $C 2/m$ (unique axis $b$, cell choice 1), $a$=5.981(2) \AA, $b$=10.354(4) \AA, $c$=6.014(2) \AA, $\alpha$=90$^o$, $\beta$=108.800(1)$^o$, $\gamma$=90$^o$. $R_f$=0.0478. $\chi^2$=1.44.  $U$ have units of \AA$^2$}
\begin{tabular}{c|c|c|c|c|c|c|c|c|c|c|c|ccc}
\hline
atom & $type$ &$site$& $x$& $y$ &  $z$  & $U_{equiv}$ & $U_{11}$ & $U_{22}$ &$U_{33}$ &$U_{23}$ &$U_{13}$ &$U_{12}$ &\\
\hline
Ru1 & Ru & $4h$&0    & 0.16653(6)  & 1/2    & 0.0041(3) & 0.0039(5)  &  0.0039(4) &  0.0046(5)  &  0      &   0.0017(3)  & 0      &\\
Cl1 & Cl & $4i$ &0.2273(4) & 0      & 0.7359(4) & 0.0049(5) & 0.0044(10) &  0.0044(8) &  0.0056(11) &  0      &   0.0012(8)  & 0      &\\
Cl2 & Cl &$8j$ &0.2504(3) & 0.17388(12) & 0.2662(3) & 0.0051(4) & 0.0049(7)  &  0.0062(7) &  0.0049(8)  &  0.0001(5)  &   0.0026(6)  & -0.0007(5) &\\
\hline
\end{tabular}
\label{strs}
\end{table}
\end{widetext}

\subsection{III-b Heat capacity }

Figure 2a shows the heat capacity of a pristine single crystal of $\alpha$-RuCl$_3$ in zero-field. Over most of the temperature range, the data can be explained by a two-dimensional (2D) Debye law as is the case with other layered materials like MoS$_2$ or graphite \cite{krum53} in which the layers are loosely bound only by van der Waals forces. In such materials, several kinds of layer stacking can be expected, giving rise to a variety of distinct polytypes, due to the pseudo hexagonal symmetry of the fundamental layer unit \cite{meng05}. In $\alpha$-RuCl$_3$, at least two types of layer stacking, ABC and ABAB type, have been proposed in the earlier report \cite{arnab15} based on two propagation vectors found in the ordered magnetic phases with $T_{N}$ = 8~K and 14~K, respectively. With the monoclinic lattice revealed by the single crystal x-ray diffraction, we illustrate these two types of stacking, ABC in Fig.~1b, and ABAB in Fig.~1c. In a perfect RuCl$_3$ lattice, one will expect almost ideal ABC-type stacking since every layer slides by nearly $a$/3 perpendicular to the honeycomb edge from the adjacent layer. This stacking sequence can be broken, for example, by the formation of stacking faults. This can easily lead to ABAB type order where the system prefers the $a$/3 layer-shift to slide back and forth (blue guide lines along $c$ in both Fig.~1b and Fig.~1c.).

The fragility of the stacking sequence in these quasi-2D crystals is evident from the heat capacity data shown in Fig.~2c. The as-grown crystal with minimal stacking faults shows one sharp magnetic transition at $T_{N}$=7~K as illustrated by heat capacity measurements in Fig.~2a, and confirmed with single crystal neutron diffraction (Fig.~2b) measured on magnetic Bragg peak (0, 1,  1/3) (discussed below).  If a typical crystal is not handled with enough care the $T_N$ loses its sharpness (Fig.~2c). If this crystal is manually deformed by bending it back-and-forth a few times, new magnetic transitions appear – most prominently the one at $T_{N}$=14~K, while the $T_{N}$ of the transition near 7~K also increases slightly to approximately 8~K, in agreement with the previous report \cite{arnab15}. This curve matches other reported susceptibility measurements \cite{majumder15, sears15}. Further deforming the crystal altogether obliterates the magnetic transition at 7~K (brown curve-3). The transition at 14~K is broad, and matches the magnetic transition in a powder sample (blue curve-4) with an ABAB type magnetic order\cite{str1,str2,kobayashi92,arnab15} as well as single crystals reported in Ref.\cite{radu15}.  Thus the material has at least two polytypes:  a quasi-trilayer and a quasi-bilayer stacking with two different N\'eel temperatures. Our as-grown crystals consistently show a pure quasi-trilayer stacking symmetry as shown in Fig.~1b. This quasi-bilayer stacking as shown in Fig.~1c is abundant in imperfect crystals or polycrystalline samples. If this type of stacking fault dominates the sample, it could be responsible for the collinear ABAB type out-of-plane antiferromagnetism observed in Refs. \cite{arnab15, radu15} below $T_{N}$ =14~K.  Given the weak van der Waals interaction orders of magnitude lower than 1 meV \cite{kim15}, it takes little energy to cause stacking faults.

\subsection{III-c Neutron diffraction }

Single crystal neutron diffraction was performed at low-temperature on a large size crystal as shown in the inset of Fig.~2a.  While all of the reflections can be indexed in a reported trigonal lattice $P3_112$,  their intensities are incompatible with a trigonal space group but conform to $C2/m$ space group consistent with the x-ray refinements. Fig.~3a shows a scan of $(1, 1, L)_T$, corresponding to (2, 0, $L$/3) in monoclinic notation \cite{matrix}, at two temperatures. The structure was refined including all three 120$^o$ twins (Fig.~3b), which propagate along the three directions perpendicular to six honeycomb edges with $a$/3 lattice shift.  They define three equivalent kinds of structural domains that may exist in a larger crystal. The mosaicity along the stacking direction is 0.5$^o$ as shown in Fig.~3c, which was limited by the instrumental resolution, and indicates that the three 120$^o$ rotated monoclinic structure domains well matched to each other. The reflections can be separated into three equivalent sets, and each set can be well fit with the untwinned monoclinic structure determined by x-ray diffraction. The fit quality is displayed in Fig.~3d by plotting the observed squared structure factors versus the calculated ones.

\begin{figure}[tbp]
\linespread{1}
\par
\begin{center}
\includegraphics[width=3.7 in]{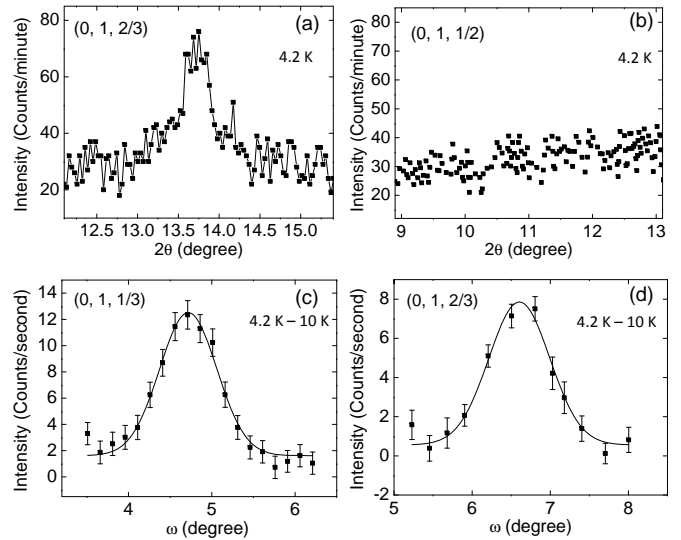}
\end{center}
\par
\caption{Single crystal neutron diffraction measured on $\alpha$-RuCl$_3$ at 4.2~K.  (a) 2$\theta$ scan at (0, 1, 2/3).  (b) The 2$\theta$ scan at (0, 1, 1/2) does not show a peak signal, i.e., no magnetic order related to the wave vector $k$=(0, 1, 1/2) exists in the measured crystal. (c) and (d) Rocking curve scans at the magnetic peak (0, 1, 1/3) and (0, 1, 2/3) indicating a magnetic order with the propagation vector $k$=(0, 1, 1/3). The magnetic signal was obtained by subtracting the corresponding scan above $T_N$ at 10~K.}
\end{figure}

The two magnetic orders identified in Ref. \cite{arnab15}, (1/2, 0, 1)$_T$ with $T_N$ = 8~K (ABC stacking) and (1/2, 0, 3/2)$_T$ with $T_N$ = 14~K (ABAB stacking), would appear as (0, 1, 1/3) and (0, 1, 1/2) type magnetic wave vectors, respectively, in the $C2/m$ space group. As expected from the heat capacity results, single crystal neutron diffraction shows a magnetic peak at $k$=(0, 1, 1/3) while no peak intensity is at $k$=(0, 1, 1/2) (Fig.~4b), ruling out the chance of complications arising from a mixture of ABC and ABAB stacking. To obtain the correct intensities, the same scan for each was performed above $T_N$ at 10~K and was used as the background. The temperature dependence of the intensity of this order was measured at the peak (0, 1, 1/3) (Fig.~2b), which is also consistent with $T_{N}$  = 7~K, and the intensity saturates below 5~K. The rocking curve scans measured at 4.2~K for the magnetic peaks (0, 1, 1/3) and (0, 1, 2/3) are shown in Fig.~4c-d. For the
magnetic structure refinement, the data were collected at 4.2~K, where the intensity has saturated, in order to obtain the correct value of the ordered moment.  Data for other magnetic peaks related to the propagation vector of $k$=(0, 1, 1/3) and -$k$ were collected in the same way as those two peaks shown in Fig.~4c-d. Each of the three structural domains produces a set of magnetic peaks characterized by $k$=(0, 1, 1/3).   If one failed to consider the presence of the structural domains then the existence of three families of propagation vectors, $k_1$=$\pm$(0, 1, 1/3), $k_2$=$\pm$(1/2, 1/2, 1/6) or (1/2, 1/2, 1/2), $k_3$=$\pm$(1/2, -1/2, -1/6) or (1/2, -1/2 ,-1/2), would be inferred. Interestingly, without considering the $L$-components, one would then obtain the same propagation vectors as reported in Na$_2$IrO$_3$, $\pm$(0, 1), $\pm$(0.5, 0.5), $\pm$(0.5, -0.5) \cite{chun15}. In both instances these are compatible with a zigzag spin structure in each honeycomb layer.  Within the resolution of the measurement, no evidence of any structural phase transition was observed across $T_N$ between 4.2~K and 10~K, consistent with the scans in Fig.~3a.

\begin{widetext}

\begin{figure}[tbp]
\linespread{1}
\par
\begin{center}
\includegraphics[width=7. in]{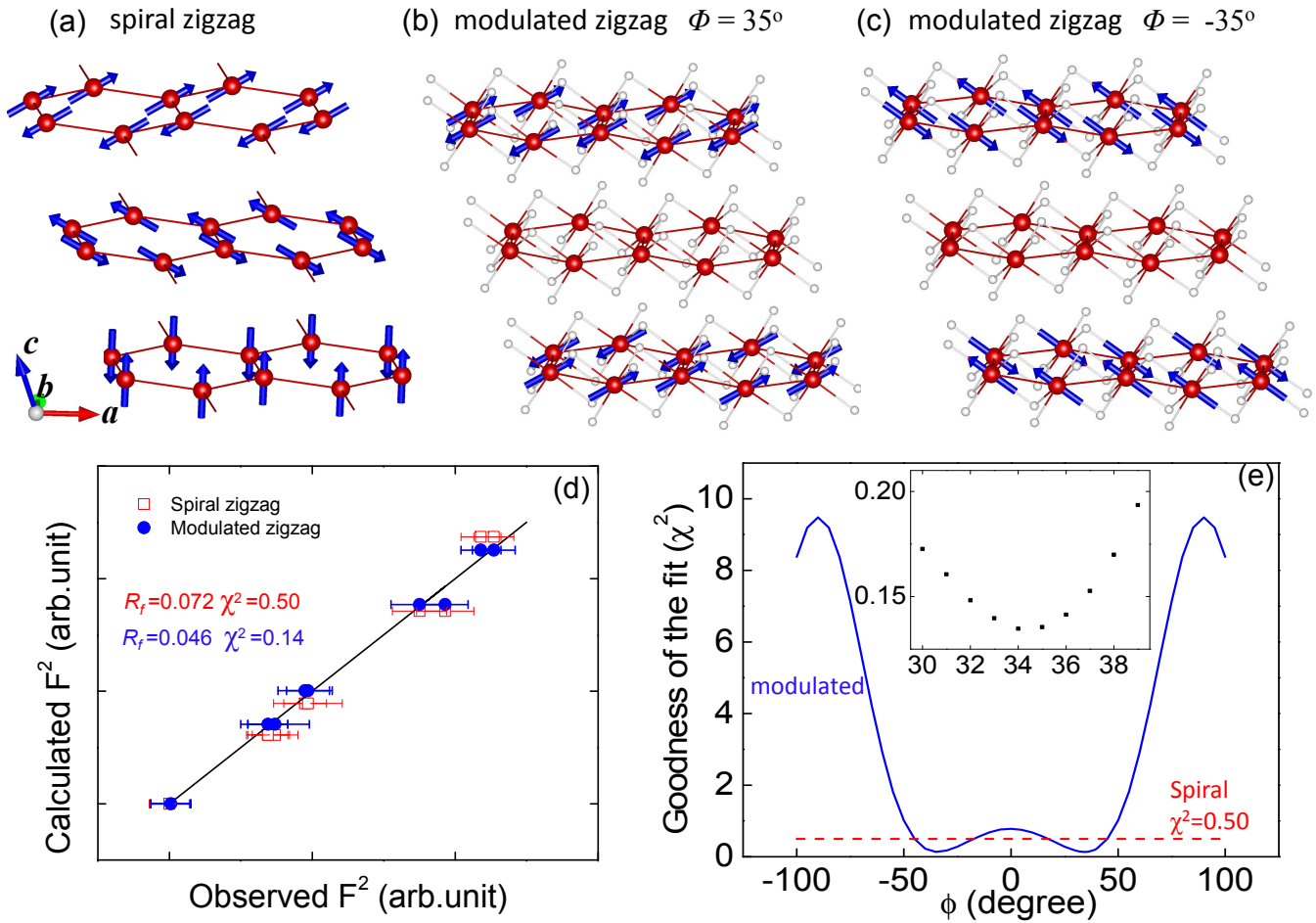}
\end{center}
\par
\caption{(Color Online) Magnetic structures of $\alpha$-RuCl$_3$ at 4.2~K. Large red balls represent Ru atoms. (a)  Spiral zigzag magnetic structure. The spins rotate 120$^o$ from one layer to the next but the spin direction in each layer is not able to be determined by the neutron diffraction data.  (b) and (c) Modulated zigzag magnetic structures showing the spins tilting $\phi$=35$^o$ and $\phi$ = -35$^o$ out of the $ab$ plane. The special case with $\Psi$=0 yields a zero-moment layer flanked by two layers with moments of magnitude 0.52 $\mu_B$.  Small white balls represent Cl atoms. (d) The observed squared magnetic structure factors versus the calculated ones for the magnetic structure refinement with the spiral (red) and modulated (blue) zigzag models. (e) the goodness of the fit $\chi^2$ versus the spin tilting angle $\phi$ out of the $ab$ plane for the spiral (red) and modulated (blue) zigzag models. The inset shows the minimum near $\phi$=35$^o$, the other minimum near $\phi$=-35$^o$ has the same $\chi^2$.}
\end{figure}
\end{widetext}

With the propagation vector $k$ = (0, 1, 1/3) and the lattice symmetry $C2/m$, two magnetic symmetries are possible, $Pm$ and $Pm'$, yielding two well-known magnetic models, zigzag and stripy, respectively. Both the symmetries were taken into consideration for refining our magnetic models. The only models producing a reasonable fit to the data show an in-plane zigzag structure with the moments confined to the $ac$ plane. A moment direction within that plane follows from symmetry requirements.    Given the in-plane zigzag structure, a reasonable fit to the data can be obtained with the out-of-plane spin structure corresponding to either a spiral or a modulated collinear spin configuration. The spiral-zigzag model (Fig.~5a) has a moment size of 0.45(5) $\mu_B$ at 4.2~K with the moment direction rotating from layer to layer by $\pm$2$\pi$/3. The goodness of fit ($\chi^2$ normalized by the degrees of freedom) is $\chi^2$ = 0.5 regardless of the overall phase angle with one choice shown in Fig.~5a. The modulated spin structures (Fig.~5b and Fig.~5c) have collinear spins but modulated sizes.  The spin amplitude on the $n^{th}$ layer $S_n =\mu_0$ $sin$($\Psi\pm2n\pi/3$), where $\Psi$ can be arbitrary and $\mu_0$ = 0.60(5) $\mu_B$ at 4.2~K. Refining the spin direction shows the best fit, with $\chi^2$ = 0.1, is obtained when the spins make an angle of $\phi$ = $\pm$35$^o$ away from the $ab$ plane.  The variation of $\chi^2$ with $\phi$  is shown in Fig.~5d.  The implications for $\phi$ = +35$^o$ (Fig 5b) and $\phi$ = -35$^o$  (Fig 5c) are discussed below. The resulting spin structure, for one exemplary case $\Psi$ = 0, is illustrated in Fig.~5b.  In this structure pairs of antiferromagnetically coupled layers are separated by a single zero-ordered-moment layer.  The present refinement favors the collinear model over the spiral model, but in the future it would be useful to carry out polarization analysis to rigorously distinguish between the two.

Within the resolution of the present measurement the ordered magnetic structure exhibits a commensurate three layer periodicity. In actuality, the low temperature $C2/m$ structure exhibts a pseudo-three layer stacking, with every 4$^{th}$ layer (top layer of Fig.~1b) displaced by $\delta$$a$ = 0.617(3) \AA, relative to a base layer (bottom layer of Fig.~1b). This corresponds to $\delta$$a$/$a$ = 0.028, suggesting that a more exact stacking occurs at 3$a$/$\delta$$a$,  approximately 107 layers. It is natural to consider whether this could lead to a magnetic structure with an incommensurate modulation along the $c$ axis, resulting in satellite reflections around the magnetic Bragg peaks.  In the present experiment no such peaks were detected.  If such peaks exist they may be extremely weak, or the periodicity of the incommensurate modulation may be too long (exceeding 200 layers) to be detected in the current measurement.   The existing data can be explained by a purely commensurate magnetic structure.

\section{Discussion} The $C2/m$, $C2/c$, $P3_112$, and more generally $P3_1$ type space groups have similar in-plane arrangements of atoms but varying out-of-plane stacking sequences. These polytypes have similar configuration energies, with energy differences of the order of 1 meV per formula unit \cite{kim15}. This is consistent with the well-known tendency for $\alpha$-RuCl$_3$ to exhibit stacking faults\cite{str2}. It is also natural to expect that the final space group manifested can depend on the starting material purity and conditions adopted for the crystal growth. With the synthesis procedures used in the present study we have found a tendency for smaller crystals to show $C2/m$ symmetry at all temperatures, while some larger crystals form in a room temperature $P3_1$ type trigonal space group.  These room temperature trigonal crystals typically show a first order phase transition at 155~K, resulting in a low temperature $C2/m$ structure.  Further systematic investigations are ongoing to elucidate the details of this behavior.  Preliminary refinements show that the in-plane structure and the local environment of the Ru$^{3+}$ ions remain largely unmodified. This suggests that the thermal history during the material growth and handling is important, and explains why various groups have identified different space groups for this material \cite{str1, str2, kim15}. On the other hand, the low-temperature structure, which determines the magnetic ground state, has always been robustly $C2/m$ in all of our crystals.

The three-dimensional magnetic ordered structure reported in various specimens of $\alpha$-RuCl$_3$ consists of stacked zig-zag ordered layers with a periodicity that is strongly dependent on the details of the samples.  Observations to date show that powders\cite{arnab15} and crystals with significant diffuse scattering from stacking faults\cite{radu15} show a broad transition near 14~K with alternating layers coupled antiferromagnetically.  The crystals reported in this study show a single sharp transition at 7~K with a three layer periodicity. The origin of this difference may be related to frustration of the inter-plane interactions as discussed below.

Previous studies of bulk properties such as the susceptibility have led to the suggestion that the inter-plane magnetic interactions are antiferromagnetic\cite{majumder15, sears15}.  These interactions are naturally satisfied if the predominant stacking sequence is dominated by repeated monolayer or bilayer sequences, which certainly will be the case if stacking faults are plentiful as may be expected, for example, in powders, where the stacking sequence could be nearly random.  The crystals measured in the present study exhibit a pseudo three layer stacking sequence with minimal stacking faults.  Diffraction alone cannot determine the phase angle $\Psi$ characterizing the exact modulation in the best fit collinear structure.   To gain insight we consider the simple cases obtained for special values of $\Psi$ when two-thirds of the layers have the same magnitude of moment. These special cases can be more generally represented by $\Psi(m)$=$\pi$$m$/6 ($m$ is an integer) with even and odd $m$, respectively. For even $m$ (e.g. $\Psi$=0), the three consecutive planes have moments (+0.52, 0, -0.52) (Fig 5b). When $m$ is odd the same moments are (-0.3, 0.6, -0.3). The latter structure, with back to back ferromagnetically stacked planes, will have a higher energy if indeed the inter-planar interactions are predominantly antiferromagnetic.   The solution with two antiferromagnetically aligned planes and one disordered makes sense physically if one considers that the   pseudo-three layer crystal structure has a tendency to frustrate the antiferromagnetic interactions and this is the solution that relieves the frustration with minimal energy cost. This possibility remains to be tested in future measurements sensitive to local moments, for example muon spin rotation, which might distinguish the various possible scenarios.

The refinement of the collinear structure shows that for any value of $\Psi$ the best fit for the spin direction points approximately 35$^o$ away from the $ab$ plane (55$^o$ away from $c^*$).   As discussed in ref.\cite{jackeli09}, for edge sharing octahedra the spin co-ordinate basis vectors make an angle of $acos$($\sqrt{2/3}$)= $\pm$35.26$^o$ to the $ab$ plane.  The two directions are not physically equivalent in the monoclinic structure, but the presence of monoclinic domains makes it impossible to distinguish between them in the present experiment.  With the conventions used in the paper, the choice of tilt angle $\phi$ = -35$^o$ (Fig.~5c) corresponds to the moment direction pointing along a Ru-Cl bond direction.  This is precisely what would be expected for a leading order antiferromagnetic Kitaev interaction with a ferromagnetic Heisenberg term\cite{chaloupka13,khaliullin05}, and, significantly, is in agreement with the conclusions derived from inelastic neutron scattering\cite{arnab15}.  The solution with $\phi$=+35$^o$ (Fig.~5b) would be close to the situation found in Na$_2$IrO$_3$, where the moments were  found to point 44.3$^o$ out of the $ab$ plane\cite{chun15}.    Future experiments on large untwinned single crystals of RuCl$_3$ might provide a definitive experimental determination of the sign of $\phi$, but the overall physical picture suggests that the moments point along the Ru-Cl ligand.

As discussed by Johnson \textit{et al.}~\cite{radu15} in the $C2/m$ structure one of the three Ru$^{3+}$-Ru$^{3+}$ distances in the honeycomb lattice is inequivalent to the others.  This small distortion may relieve some frustration and contribute to the formation of the low temperature zigzag magnetic order with a small ordered moment.  Our refinements show that the magnitude of this distortion is less than 0.2\%. As noted previously\cite{arnab15} this can result in an anisotropic Kitaev term, equivalent to an additional uniaxial Ising term, in the effective spin Hamiltonian. This is one of several possible terms that can result in a spin gap within a classical spin wave model of magnetic excitations, although such a gap is also expected to arise from quantum fluctuations which will be strong in $\alpha$-RuCl$_3$.  We believe
 it is premature to fully attribute this gap to bond anisotropy as suggested in reference\cite{radu15}.

In a pure Kitaev model magnetic order is absent and the T=0 ground state is a QSL. In $\alpha$-RuCl$_3$, a low-temperature magnetic order is believed to be formed due to sub-leading magnetic interactions in addition to a dominant Kitaev term. The in-plane order is always characteristic of the zig-zag phase.  Conversely, the precise three-dimensional magnetic order is fragile and easily altered by the formation of low-energy stacking faults \cite{kim15}, as proved by the drastic changes of $T_N$ and $k$. From the high temperature magnetic susceptibility, the effective paramagentic moment has been estimated as roughly $g\sqrt{(S(S+1))}$=2.2$\mu_B$ \cite{majumder15,kobayashi92,arnab15}. Assuming $g$=2, a fully ordered moment projected onto one axis would yield approximately 1.4 $\mu_B$ in a neutron diffraction experiment. The measured ordered moment of 0.45 $\mu_B$ is therefore 1/3 of the fully ordered Ru$^{3+}$ moment. With this low ordered moment one expects that quantum spin fluctuations are dominant, and the full spectrum of magnetic excitations will not be well-described by a classical spin-wave picture. One approach might be to consider the existence of a small ordered moment as a perturbation to a QSL ground state \cite{arnab15}.

\section{Conclusions} Single crystals of $\alpha$-RuCl$_3$ with minimal stacking faults are seen to consistently exhibit a low temperature crystal structure described by the $C2/m$ space group.  The magnetic structure consists of zig-zag planes stacked with a three layer periodicity.  The best refinement of the magnetic structure is collinear with the moment directions oriented $\pm$35$^o$ from the $ab$ plane, consistent with expectations for edge-sharing octahedra.  We argue that a structure with two antiferromagnetically aligned layers and one disordered layer is consistent with the refinements and is likely on physical grounds.

\section{Acknowledgments}

\begin{acknowledgments}
We thank Feng Ye, Adam Aczel, and Johannes Knolle for discussions during various stages of this research, and Songxue Chi for helping with the experiment setup at HB-3A. We have benefited from discussions with Hae-Young Kee, Giniyat Khaliullin, and George Jackeli.  The work at ORNL HFIR was sponsored by the Scientific User Facilities Division, Office of Science, Basic Energy Sciences, U.S. Department of Energy. JQY and CAB were supported by the U. S. Department of Energy, Office of Science, Basic Energy Sciences, Materials Sciences and Engineering Division. DGM was supported by the Gordon and Betty Moore Foundation’s EPiQS Initiative through Grant GBMF4416.

HC and AB contributed equally to this work.
\end{acknowledgments}


\begin{thebibliography}{99}
\bibitem{kitaev06} A. Kitaev, Ann. Phys. (N.Y.) 321, 2 (2006).
\bibitem{nayak08} C. Nayak et~al., Rev. Mod. Phys. 80, 1083-1159 (2008).
\bibitem{yamashita10} M. Yamashita et~al., Science 328, 1246-1248 (2010).
\bibitem{han12} T.-H. Han et~al., Nature  492, 406-410 (2012).
\bibitem{knolle14} J. Knolle, D.L. Kovrizhin, J.T. Chalker, and R. Moessner, Phys. Rev. Lett. 112, 207203 (2014). 
\bibitem{jackeli09} G. Jackeli and G. Khaliullin, Phys. Rev. Lett. 102, 017205 (2009). 
\bibitem{singh10} Y. Singh and P. Gegenwart, Phys. Rev. B 82, 064412 (2010).
\bibitem{biffin14} A. Biffin, R. D. Johnson, S. Choi, F. Freund, S. Manni, A. Bombardi,  P. Manuel, P. Gegenwart, and R. Coldea, Phys. Rev. B 90, 205116 (2014)
\bibitem{rau15} J.G. Rau, E.K. Lee and H.-Y. Kee, "Spin-Orbit Physics Giving Rise to Novel Phases in Correlated Systems$:$ Iridates and Related Materials", arXiv:1507.06323 (2015).
\bibitem{chun15}  S.H. Chun, J.-W. Kim, J. Kim, H. Zheng, C.C. Stoumpos, C.D. Malliakas, J. F. Mitchell, K. Mehlawat, Y. Singh,	Y. Choi, T. Gog, A. Al-Zein, M.M. Sala, M. Krisch, J. Chaloupka, G. Jackeli, G. Khaliullin and B.J. Kim,  Nature Physics 11, 462 (2015). 
\bibitem{singh12} Y. Singh, S. Manni, J. Reuther, T. Berlijn, R. Thomale, W. Ku, S. Trebst, and P. Gegenwart, Phys. Rev. Lett. 108, 127203 (2012). 
\bibitem{chaloupka13} J. Chaloupka, G. Jackeli, and G. Khaliullin, Phys. Rev. Lett. 110, 097204 (2013).
\bibitem{coldea14} A. Biffin, R. D. Johnson, I. Kimchi, R. Morris, A. Bombardi, J.G. Analytis, A. Vishwanath, and R. Coldea, Phys. Rev. Lett. 113, 197201 (2014).
\bibitem{choi12} S.K. Choi, R. Coldea, A.N. Kolmogorov, T. Lancaster, I.I. Mazin, S.J. Blundell, P.G. Radaelli, Y. Singh, P. Gegenwart, K.R. Choi, S.-W. Cheong, P.J. Baker, C. Stock, and J. Taylor, Phys. Rev. Lett. 108, 127204 (2012).
\bibitem{ye12} F. Ye, S. Chi, H.B. Cao, B.C. Chakoumakos, J.A. Fernandez-Baca, R. Custelcean, T.F. Qi, O.B. Korneta, and G. Cao, Phys. Rev. B 85, 180403(R) (2012). 
\bibitem{sears15}J. A. Sears, M. Songvilay, K. W. Plumb, J. P. Clancy, Y. Qiu, Y. Zhao, D. Parshall, and Y.-J. Kim, Phys. Rev. B 91, 144420 (2015). 
\bibitem{sandilands15} L.J. Sandilands, Y. Tian, K.W. Plumb, Y.-J. Kim, and K.S. Burch, Phys. Rev. Lett. 114, 147201 (2015). 
\bibitem{majumder15} M. Majumder, M. Schmidt, H. Rosner, A. A. Tsirlin, H. Yasuoka, and M. Baenitz, Phys. Rev. B 91, 180401(R) (2015). 
\bibitem{kobayashi92} Y. Kobayashi, T. Okada, K. Asai, M. Katada, H. Sano and F. Ambe, Inorg. Chem. 31, 4570-4574 (1992). 
\bibitem{radu15} R.D. Johnson, S.C. Williams, A. A. Haghighirad, J. Singleton, V. Zapf, P. Manuel, I.I. Mazin, Y. Li,  H.O. Jeschke,  R. Valent\'i, and R. Coldea, Phys. Rev. B 92, 235119 (2015).
\bibitem{arnab15} A. Banerjee, C.A. Bridges, J-Q. Yan, A.A. Aczel, L. Li, M.B. Stone, G.E. Granroth, M.D. Lumsden, Y. Yiu, J. Knolle, D.L. Kovrizhin, S. Bhattacharjee, R. Moessner, D.A. Tennant, D.G. Mandrus, and S.E. Nagler, "Proximate Kitaev Quantum Spin Liquid Behavior in $\alpha$-RuCl$_3$", arxiv.org/pdf/1504.08037 (2015). (To appear in Nature Materials).
\bibitem{kubota15} Y. Kubota, H. Tanaka, T. Ono, Y. Narumi, and K. Kindo,  Phys. Rev. B 91, 094422 (2015). 
\bibitem{str1} J.M. Fletcher, W.E. Gardner, E.W. Hooper, K.R. Hude, F.H. Moore, and J.L. Woodhead, Nature 199, 1089-1090 (1963).
\bibitem{str2} J.M. Fletcher, W.E. Gardner, A.C. Fox, G. Topping, J. Chem. Soc. A, 1038-1045 (1967).
\bibitem{chaloupka10} J. Chaloupka, G. Jackeli, and G. Khaliullin, Phys. Rev. Lett. 105, 027204 (2010). 
\bibitem{rau14} J.G. Rau, E.K.-H. Lee and H.-Y. Kee, Phys. Rev. Lett. 112, 077204 (2014). 
\bibitem{chaloupka15} J. Chaloupka and G. Khaliullin, Phys. Rev. B 92, 024413 (2015). 
\bibitem{rous15} I. Rousochatzakis, J. Reuther, R. Thomale, S. Rachel and N. B. Perkins, arXiv:1506.09185 (2015). 
\bibitem{kim15} H.-S. Kim and H.-Y. Kee, "Structural properties of $\alpha$-RuCl$_3$$:$ an ab-initio study", arXiv:1509.04723 (2015).
\bibitem{smithyoder56} J.V. Smith and H.S. Yoder Jr., Mineral. Mag. 31, 209235 (1956). 
\bibitem{str0} E. V. Stroganov and K. V. Ovchinnikov, Ser. Fiz. i Khim.12, 152 (1957).
\bibitem{strmono} K. Brodersen, G. Thiele, H. Ohnsorge, I. Recke, and F. Moers, J. Less Common Metals 15, 347 (1968).
\bibitem{matrix} The transformation matrix [1/2, -1/2, 0; 1/2, 1/2, 0; 1, 0, 3] can be used as a guide to transform the reflection indices in the monoclinic cell $C2/m$ (M) to those in the trigonal $P3_112$ (T) cell.  Thus for example, the peaks (0, 0, 1)$_M$,  (0, 2, 0)$_M$,  (0, 1, 1/3)$_M$, and  (1/2, 1/2, 1/6)$_M$ will transform to (0, 0, 3)$_T$, (-1, 1, 0)$_T$, (-1/2, 1/2, 1)$_T$, and (0, 1/2, 1)$_T$ respectively, and vice-versa.
\bibitem{hyde65} K. R. Hyde, E. W. Hooper, J. Waters and J. M. Fletcher, J. Less-Common Metals, 8 , 428 (1965). 
\bibitem{max3d} J. Britten and W. Guan, MAX3D - A Program for the Visualization of Reciprocal Space. IUCr Commission on Crystallographic Computing Newsletter No. 8, 96-108 (2007).
\bibitem{rigaku05} Rigaku (2005) CrystalClear. Rigaku Corporation, Tokyo, Japan.
\bibitem{higashi} Higashi, T. (2000) ABSCOR. Rigaku Corporation, Tokyo, Japan.
\bibitem{sheldrick} G. M. Sheldrick, Acta Crystallogr. A64, 112-122 (2008);
\bibitem{farrugia} J. Farrugia, J. Appl. Cryst. 45, 849-854 (2012). ;
\bibitem{burla} M.C. Burla, R. Caliandro, M. Camalli, B. Carrozzini, G.L. Cascarano, C. Giacovazzo, M. Mallamo, A. Mazzone, G. Polidori, and R. Spagna, J. Appl. Cryst. 45, 351-356 (2012).
\bibitem{momma} K. Momma and F. Izumi, J. Appl. Crystallogr. 41, 653-658 (2008).
\bibitem{hb3a} B.C. Chakoumakos, H.B. Cao, F. Ye, A.D. Stoica, M. Popovici, M. Sundaram, W. Zhou, J.S. Hicks, G.W. Lynn and R.A. Riedel, J. Applied Cryst., 44, 655 (2011).
\bibitem{fullprof} J. Rodriguez-Carvajal, Physica B 192 55 (1993).
\bibitem{krum53} J. Krumhansl and H. Brooks, J. Chem. Phys. 21, 1663(1953). 
\bibitem{meng05} Y. S. Meng, G. Ceder, C. P. Grey, W.-S. Yoon, M. Jiang, J. Br\'eger, and Y. Shao-Horn, Chem. Mater., 17, 2386 (2005).
\bibitem{khaliullin05} G. Khaliullin, Prog. Theor. Phys. Suppl. 7, 160, 155 (2005).


\end{thebibliography}
\end{document}